\documentclass[prd,twocolumn,showpacs,preprintnumbers,amsmath,amssymb]{revtex4}  
\usepackage{epsfig}%
\usepackage{longtable}
\begin{document}
\setlength{\topmargin}{-0.25in}
\preprint{U. of IOWA Preprint}

\title{An example of optimal field cut in lattice gauge perturbation theory}


\author{L.Li}
\author{Y. Meurice}
 \altaffiliation[Also at ]{Obermann Center for Advanced study}%Lines break
\email[]{ yannick-meurice@uiowa.edu}
%\thanks{This work was supported in part by the DOE}
\affiliation{Department of Physics and Astronomy\\ The University of Iowa\\
Iowa City, Iowa 52242 \\ USA
}


\date{\today}

\begin{abstract}
We discuss the weak coupling expansion of a one plaquette $SU(2)$ lattice gauge theory. We show that the conventional perturbative series for the partition function has a zero radius of convergence and is asymptotic. 
The average plaquette is discontinuous at $g^2=0$. 
However, the fact that $SU(2)$ is compact provides a perturbative 
sum that converges toward the correct answer for positive $g^2$. This alternate method amounts to introducing a specific coupling dependent field cut, that turns the coefficients into  
$g$-dependent quantities. Generalizing to an arbitrary field
cut, we obtain a regular power series with a finite radius of convergence. At any order in the modified perturbative procedure, and for a given coupling, it is possible 
to find at least one (and sometimes two) values of the field cut that provide 
the exact answer.  This optimal field cut can be determined 
approximately using the strong coupling expansion. This allows us to interpolate 
accurately 
between the weak and strong coupling regions. We discuss the extension of the method to 
lattice gauge theory on a $D$-dimensional cubic lattice.

\end{abstract}

%\pacs{05.50.+q, 11.10.Hi, 64.60.Ak, 75.40.Cx}
%\keywords{}

\maketitle

\section{Introduction}
Lattice gauge theory incorporates 
essential features of the strong interactions at short distance (asymptotic freedom) 
and large distance (confinement). 
Expansions in $1/\beta=g^2/2N$ and $\beta$ usually provide good approximations for the average value of gauge invariant quantities in the limit of small or 
large $\beta$. However, calculations in the intermediate region often 
require a numerical approach. 

There exists a  general method for calculating Wilson' s or Polyakov's 
loops in powers of $1/\beta$ 
\cite{heller84} in pure $SU(N)$ gauge theories (see Ref. \cite{capitani02} for
a more complete set of references on lattice perturbation theory). 

\begin{figure}[ht]
\includegraphics[width=3.4in,angle=0]{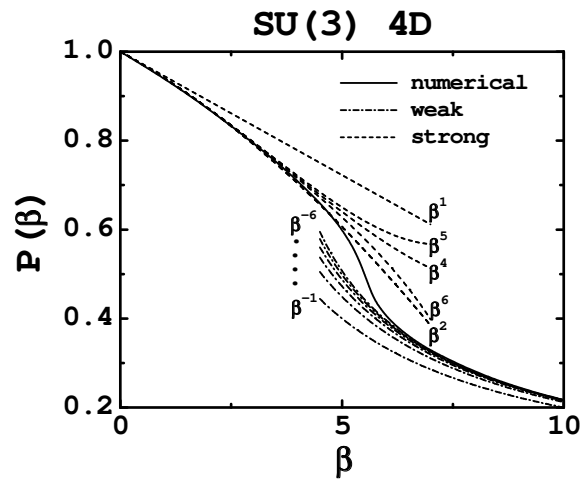}
\caption{$P$ versus $\beta$ for $SU(3)$ in 4 dimensions. The solid line represents 
the numerical values; the dashed lines on the left, successive orders in the strong 
coupling expansion; the dot-dash lines on the right, successive orders in  the 
weak coupling expansion.
\label{fig:su3}}
\end{figure}

Much effort has been devoted calculating 
\begin{equation}
P\equiv \left\langle(1/N_p)\sum_p (1-(1/N)ReTrU_p)\right\rangle
\end{equation}
where $U_p$ denotes the usual product of links along 
a $1\times1$ plaquette and $N_p$ the number of plaquettes. $P$ can be obtained by  taking the 
derivative with respect to $\beta$ of the free energy density. 
Exact calculations of the coefficients of $P$ up to order 3 in $1/\beta$
\cite{alles93} and numerical calculations at order 8 \cite{direnzo95} and 10 \cite{direnzo2000} are available. 
The accuracy of the weak and strong coupling expansions at successive orders is shown in Fig. \ref{fig:su3} for $SU(3)$ in 4 dimensions. The figure makes clear that 
in the region $5<\beta<6$, none of the expansions (in powers of $\beta$ or $1/\beta$) is accurate.  
Unfortunately, this region is precisely the ``scaling window'' where one can extract 
information relevant to the continuum limit. In addition, the convergence of the 
weak coupling expansion is not completely understood. The analysis of the numerical series 
\cite{rakow2002} may suggest the possibility of a finite radius of convergence (the center of the circle being $g^2=0$).
This possibility is not expected on general grounds and is  
in contradiction with the discontinuity of $P$ when $g^2$ changes sign 
\cite{gluodyn04,anlqcd}. 
We are not aware of any independent argument in favor of a finite 
radius of convergence, and the most likely outcome is that the factorial 
growth of the series takes over at higher order. In Ref. 6, this order is 
estimated to be approximately 25, which is out of reach of numerical 
calculations.

In this article, we show that for a $SU(2)$ lattice gauge model on 
a single plaquette, the weak 
coupling expansion is asymptotic (has zero radius of convergence), 
but that it is possible 
to modify the perturbative procedure in order to get a convergent 
sum, which is an ``expansion'' in powers of $1/\beta$ but 
with $\beta$-dependent coefficients, that 
is accurate even in the strong coupling region. This work is motivated by 
recent results obtained in the case of scalar field theory \cite{convpert,optim03}
where the answer for similar questions in the case of nontrivial models 
can be guessed correctly by 
considering a single site integral. This is briefly reviewed in Sec. 
\ref{sec:motivations}. The main point is that the large order behavior of 
perturbation theory is related to large field configurations and that by 
cutting-off these configurations appropriately, we can obtain a series 
that converges to a value exponentially close to the exact one 
(for instance, errors of order ${\rm e}^{-\lambda\phi^4_{max}}$ for the simple 
integral discussed in Ref. \cite{optim03}).
Hereafter, we follow the same path for lattice gauge theories.

The $SU(2)$ model is introduced in Sec. \ref{sec:model} where we also discuss its 
connection to Bessel functions. It is worth noting that for 
compact groups, there are no large field contributions and consequently, the 
factorial growth of the perturbative series comes from adding the tails of integration, 
as done in asymptotic analysis of integrals \cite{bleistein} and in the conventional procedure used in lattice perturbation theory\cite{heller84}. This is explained in Section 
\ref{sec:weak}. In many quantum problems, the lack of convergence can be traced 
to the behavior of the model at negative coupling (see, however, Ref. \cite{bender98} 
for a proper definition). This question has been discussed for lattice gauge models in 4 dimensions \cite{gluodyn04}. In Sec. \ref{sec:neg}, we argue that there should be an essential 
singularity at zero coupling for the one plaquette model. In Sec. \ref{sec:instanton}, 
we show that the regular perturbation series (with the integration tails added) misses 
``instanton effects'' of the form 
$\beta^{-1}{\rm e}^{-2\beta}$. 

\begin{figure}[ht]
\includegraphics[width=3.4in,angle=0]{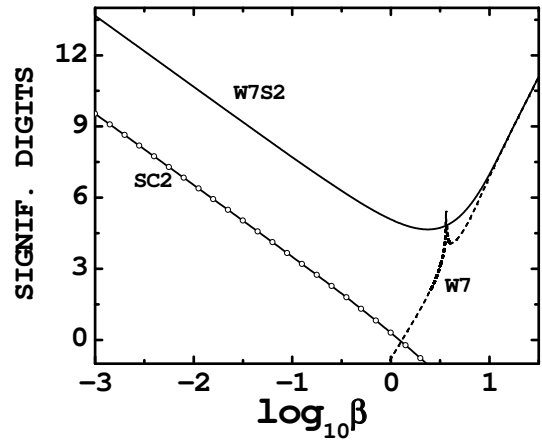}
\caption{Number of significant digits for the $SU(2)$ one plaquette integral, at 
order 7 in the weak coupling (dotted line W7), at order 2 in the strong coupling (empty circles SC2) and at order 7 of the modified perturbative method proposed here with 
an optimal field cut determined pointwise using the strong coupling expansion at order 2 (solid line W7S2).   
\label{fig:pico}}
\end{figure}

In Sec. \ref{sec:cut}, we propose to modify the conventional perturbative method by 
introducing a field cut. With this modification, the series converges toward a 
value which in general is different than the exact one. However, at a given order in 
the weak coupling expansion, it is possible to pick an 
optimal field cut such that for a given coupling the answer is exact.
For the integral studied in this article, we found at least one solution 
at every order. This is not necessarily the case in general. For the 
integral studied in Ref. \cite{optim03}, we were able to prove that no such 
solution exists at odd order and that we could only minimize the error in that case.
In Sec. \ref{sec:optim}, we use the strong coupling expansion to determine approximately
this optimal field cut. In this approach, the field cut is given as a power series 
in $\beta$. A numerical study indicates that this series has a finite radius of 
convergence which increases with the order in $1/\beta$ considered. 
The method that we propose allows us to interpolate between the weak and strong 
coupling region. This is depicted in Fig. \ref{fig:pico}, which is the prototype 
of what we expect to accomplish in general. In the conclusions, we consider the implementation of the method for 
$D$-dimensional models and discuss three practical ways to calculate the 
modified coefficients.

\section{Motivations}
\label{sec:motivations}

A common challenge for quantum field theorists consists in 
finding accurate methods in regimes where existing expansions break down. 
In the renormalization group language, this amounts to finding acceptable interpolations for the flows in intermediate regions between fixed points.
A discussion of this question for lattice gauge theories 
can be found in Refs. \cite{kogut79,kogut80}. 
In the case of scalar field theory, the weak coupling expansion is unable to reproduce the exponential suppression of the large field configurations operating 
at strong coupling. This problem can be cured \cite{convpert} by 
introducing a large field cutoff $\phi _{max}$ which eliminates Dyson's instability.
One is then considering a slightly different problem, however a judicious choice of $\phi _{max}$ can be used to reduce or eliminate \cite{optim03} the discrepancy with the original problem (i. e., the problem with no field cutoff). 
This optimization procedure can be approximately 
performed using the strong coupling expansion and naturally bridges the gap between 
the weak and strong coupling expansions. 

The study of the simple integral 
\begin{equation}
\int_{-\infty}^{+\infty}d\phi e^{-\frac{1}{2}\phi^2-\lambda \phi^4}\neq \sum_0^{\infty}
\frac{(-\lambda)^l}{l!} \int_{-\infty}^{+\infty}d\phi e^{-\frac{1}{2}\phi^2}\phi^{4l}
\label{eq:int}
\end{equation}
provides a good understanding about the role of large field configurations in 
the perturbative series. It helps identifying general features of the the effect 
of a field cut.
In particular, the dependence of the accuracy of the modified perturbative series
on the coupling and the field cut, is qualitatively very similar for the 
integral, the anharmonic oscillator and the hierarchical model in 3 dimensions
(see the similarity among the three parts of Fig. 2 in \cite{convpert}).

\begin{figure}[ht]
\includegraphics[width=3.4in,angle=0]{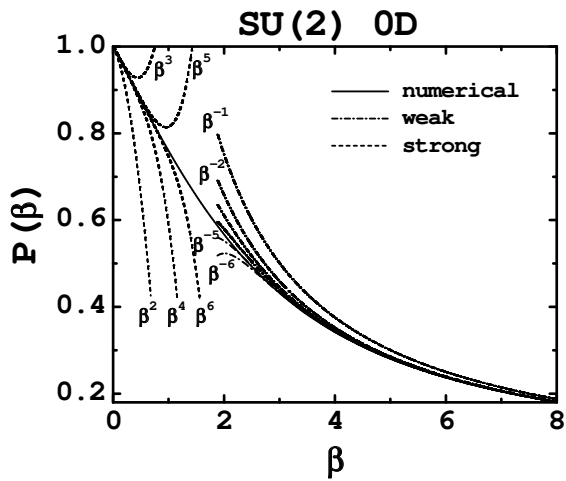}
\caption{ $P$ versus $\beta$ for $SU(2)$ on one plaquette. The solid line represents 
the numerical values; the dashed lines on the left, successive orders in the strong 
coupling expansion; the dot-dash lines on the right, successive order in  the 
weak coupling expansion.
\label{fig:su2one}}
\end{figure}

In order to generalize this procedure to gauge theory, we will first consider the simplest 
possible gauge model, namely the one plaquette integral
\begin{equation}
Z(\beta,N)=\int \prod_{l\in p} dU_l {\rm e} ^{-\beta(1-\frac{1}{N}Re TrU_p)}\ ,
\end{equation}
%with the usual convention
%\begin{equation}
%\beta=2N/g^2
%\end{equation}
After fixing the gauge so that $U=1$ on three sides of the plaquette, $Z$ becomes an integral over a single 
link
\begin{equation}
Z(\beta,N)=\int dU {\rm e} ^{-\beta(1-\frac{1}{N}Re TrU)}
\end{equation}
For an arbitrary gauge fixing prescription, $TrU$ becomes $Tr(U(g)U)$ with $g$ arbitrary 
and $Z$ is $g$-independent by virtue of the invariance of the Haar measure $dU$.
This integral and its moments appear in the strong coupling expansion 
\cite{kogut79,munster80,falcioni80,smit82}
and in the mean field treatment \cite{drouffe83} of $SU(N)$ gauge theories. In the one plaquette model,
\begin{equation}
P =-\frac{d}{d\beta}{\rm ln}Z\ .
\end{equation} 
The accuracy of successive orders in the $\beta$ and $1/\beta$ described in the 
following sections is shown in Fig. \ref{fig:su2one} that can be compared with Fig. \ref{fig:su3}.
\section{The model considered here}
\label{sec:model}

In the following, we specialize the discussion to the case $N=2$ for which the Haar measure is very simple. From now on, the reference to $N$ will be dropped and we will use the notation $Z(\beta)$ for $Z(\beta,2)$. The explicit form is:
\begin{equation}
	Z(\beta)=\frac{1}{\pi}\int_0^{2\pi}d\omega \sin^2(\omega/2){\rm e} ^{-\beta(1-\cos(\omega/2))}\ .
	\label{eq:zsu2}
\end{equation}
Setting $u={\rm cos}(\omega/2)$, 
\begin{equation}
	Z(\beta)=\frac{2}{\pi}\int_{-1}^1du \sqrt{1-u^2}{\rm e}^{-\beta(1-u)}
	\label{eq:bform}
\end{equation}
and one recognizes from Eq. (8.431) of Ref. \cite{integrals} that the integral can be expressed in terms of the modified Bessel function $I_1$:
\begin{equation}
Z(\beta)=2 {\rm e}^{-\beta}I_1(\beta)/\beta
\end{equation}
Using the Taylor expansion Eq. (8.445) in Ref. \cite{integrals}, we can write
\begin{equation}
2 I_1(\beta)/\beta=\sum_{l=0}^{\infty} \frac{1}{l!\Gamma(l+2)}(\beta/2)^{2l} \ .
\end{equation}
As in the case of the integral of Eq. (\ref{eq:int}), the presence of the factorial at the denominator implies that the strong coupling expansion 
(in powers of $\beta =4/g^2$) converges over the entire complex plane.

\section{The weak coupling expansion}
\label{sec:weak}

Assuming $\beta>0$, we set $t=\beta(1-u)$ in Eq. (\ref{eq:bform}) yields 
\begin{equation}
	Z(\beta)=(2/\beta)^{3/2}\frac{1}{\pi}\int_0^{2\beta}dt t^{1/2}
	{\rm e}^{-t}\sqrt{1-(t/2\beta)}
	\label{eq:tint}
\end{equation}
If we expand the square root in the integral and exchange the sum and the integral (the validity of this procedure will be discusssed in Sec. \ref{sec:cut}), 
we obtain a converging sum:
\begin{equation}
\label{eq:notail}
	Z(\beta)=(\beta\pi)^{-3/2} 2^{1/2} 
	\sum_{l=0}^{\infty} A_l(2\beta)\beta^{-l}\ ,
	\end{equation}
	with 
	\begin{equation}
	A_l(x)\equiv 2^{-l}
	\frac{\Gamma(l+1/2)}{l!(1/2-l)}\int_0^{x}dt {\rm e}^{-t}t^{l+1/2}\ ,
	\label{eq:al}
\end{equation}
 
The convergence of the 
sum in Eq. (\ref{eq:notail}) can be established from the bounds 
\begin{equation}
\frac{{\rm e}^{-2\beta}}{l+3/2}(2\beta)^{l+3/2}	<\int_0^{2\beta}dt {\rm e}^{-t}t^{l+1/2} < \frac{1}{l+3/2}(2\beta)^{l+3/2}\ ,
\end{equation}
and the fact that $\Gamma(l+1/2)/l! <1$ for $l\geq1$. (This is true for $l=1$ and can be proved by induction multiplying the inequality by $(l+1/2)/(l+1) <1$).
Consequently, the sum converges at the same rate as $\sum l^{-2}$. 
Note that as in the case of the 
ground state  of the quantum mechanical double well, the first term is positive but all the remaining terms are negative.

Obviously, Eq. (\ref{eq:notail}) is not a power series in $\beta^{-1}$ since the ``coefficients'' $A_l(2\beta)$ are $\beta$-dependent. 
We can now obtain the conventional asymptotic expansion by two different methods. The first one consists in adding the tails to the integrals in Eq. (\ref{eq:al}), or in other words by replacing the incomplete gamma function by the gamma function. 
This is a standard procedure in asymptotic expansions of integrals \cite{bleistein}.

On then obtain the asymptotic expansion 
\begin{eqnarray}
&\ &	Z(\beta)\sim(\beta\pi)^{-3/2} 2^{1/2} \times \nonumber \\ 
&\ &	\sum_{l=0}^{\infty} (2\beta)^{-l}
	\frac{(\Gamma(l+1/2))^2(l+1/2)}{l!(1/2-l)},
	\label{eq:tail}
\end{eqnarray}
The terms of this sum now grow like $l!/2^l$ and the series is asymptotic. 
As all the signs are negative for $l\geq 1$, the Borel transform has singularities 
on the positive real axis.

It is instructive to rederive the expansion of Eq. (\ref{eq:tail}) by following the steps of lattice 
perturbation theory \cite{heller84}. We first set $\omega =gA$ in Eq. (\ref{eq:zsu2}) and expand the action and the Haar measure in powers of $g$. This leaves us with 
the integral of a power series in $g$ over the range 0 to $2\pi/g$ for $A$. 
The asymptotic
series (\ref{eq:tail}) is then recovered by letting the range of integration 
go to infinity. As the two methods amount to calculate the coefficients 
with different variables of 
integration, we obtain the same series, as can be checked explicitly up to 
high order. We emphasize that in lattice gauge theory with compact groups, 
there are no large field contributions. It is only for practical reasons that 
the tails of integration are added. In the one plaquette example, calculating 
$A_l(2 \beta)$ instead of $A_l(\infty)$ is a very minor problem, however, this is 
a technical challenge in the case on a $D$-dimensional lattice

\section{Behavior at negative $\beta$}
\label{sec:neg}

From the integral representation Eq. (\ref{eq:bform}), the change $\beta \rightarrow -\beta$ can be made by changing $u\rightarrow-u$ and multiplying by ${\rm e}^{2\beta}$. This implies, 
\begin{equation}
Z(-\beta)={\rm e}^{2\beta} Z(\beta)\ ,
\label{eq:minbet}
\end{equation}
and 
\begin{equation}
P(\beta)+P(-\beta)=2	
\label{eq:sumrule}
\end{equation}
A similar equation \cite{gluodyn04} can be found for a $SU(2)$ pure gauge model on a cubic lattice. Since $lim_{\beta\rightarrow+\infty}P(\beta)=0$,
the limits $g^2\rightarrow 0^{\pm}$ differ by 2 and a converging 
series in $g$ about 0 is impossible.

The discontinuity in the values of $P$ near $g^2=0$ appears in a simpler model 
where the integration over $SU(2)$ is replaced by a sum over the two elements of its 
center:
\begin{equation}
	Z_{center}=\sum_{U=\pm \openone}{\rm e} ^{-\beta(1-\frac{1}{2}Re TrU)}=1+{\rm e}^{-2\beta}\ .
	\label{eq:zcenter}
\end{equation}
This implies 
\begin{equation}
P_{center}=\frac{2}{1+{\rm e}^{2\beta}}\ .
\end{equation}
The center model satisfies Eqs. (\ref{eq:minbet}) and (\ref{eq:sumrule}).
Note that Eq. (\ref{eq:zcenter}) makes clear that $Z_{center}$ has an essential singularity
at $g=0$. The asymptotic behavior of $P_{center}$ at large $|\beta|$ in the complex plane is $2(1-{\rm e}^{2\beta}+\dots)$ if $Re\beta <0$, and $2{\rm e}^{-2\beta}+\dots$, if $Re\beta >0$, with Stokes lines running along the imaginary axis.

This simplified example makes clear that the usual perturbation series is obtained by making modifications of order ${\rm e}^{-2\beta}$ (the effect of the tails of integration). 
We now proceed to estimate the order ${\rm e}^{-2\beta}$ corrections to 
the integral over the whole $SU(2)$.

\section{Accuracy of regular perturbation theory}
\label{sec:instanton}

In the study of scalar models \cite{convpert}, we have shown that if we plot the 
accuracy of perturbative series at successive orders, an envelope 
setting the
boundary of the accuracy that can be reached at any order using the usual perturbation theory 
appears. In the case of the quantum mechanical double-well, this envelope coincides 
very precisely with the instanton effect.
We expect the limitation in accuracy to be of the general form 
$g^A{\rm e}^{-B/g^2}$. For the model considered here, 
the limitation of accuracy has this generic form and we will see that 
the effect is of order $\beta^{-1}{\rm e}^{-2\beta}$. 

\begin{figure}[ht]
\includegraphics[width=3.4in,angle=0]{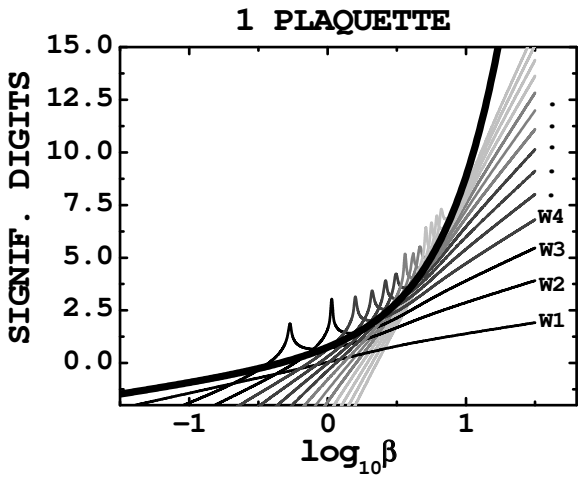}
\caption{Number of correct significant digits as a function of $\beta$ at successive orders
of the regular perturbative series Eq. (\ref{eq:tail}) for $Z(\beta)$. As the order increases from 1 to 15, the 
curves ($W1,\ W2,\dots)$ get lighter. The thick solid line is ${\rm log}_{10}(\beta^{-1}{\rm e}^{-2\beta}/Z)$.
\label{fig:envelope}}
\end{figure}

For $\beta$ not too small, the low orders the asymptotic series Eq. (\ref{eq:tail}) overestimate $Z$. As we let the order increase, the series 
crosses the true value and then start to grossly underestimate the true value.
At each order, there is a special value of $\beta$ for which the truncated 
series coincides with the exact 
answer. This explains the ``spikes'' seen  in Fig. \ref{fig:envelope}.

If we assume that for a particular value of $\beta$, the converging sum, Eq. (\ref{eq:notail}) with the integrals running from 0 to $2\beta$,  truncated at order $K$ is a good approximation of $Z(\beta)$, then the error $\delta Z(\beta, K)$ made by using the regular perturbative series, 
Eq. (\ref{eq:tail}) with the integrals running from 0 to $\infty$, truncated at the same order, is in good approximation
\begin{eqnarray}
	\delta Z(\beta, K)\simeq (\beta \pi)^{-3/2} 2^{1/2} &\times & \nonumber \\ 
	\sum_{l=0}^{K} (2\beta)^{-l}
	\frac{\Gamma(l+1/2)}{l!(1/2-l)}\int_{2\beta}^{\infty}&dt& {\rm e}^{-t}t^{l+1/2}\ ,
	\label{eq:delz}
\end{eqnarray}

Integrating by parts, dropping terms of order $\beta^{-1}$ and summing the 
resulting series, we obtain 
 
\begin{equation}
	\delta Z(\beta, K)\approx A_K{\rm e}^{-2\beta}\beta^{-1}2\pi^{-3/2}\ , 
\end{equation}
with 
\begin{equation}
A_K=-\sum_{l=0}^{K}\frac{\Gamma(l-1/2)}{l!}\ .
\end{equation}
The coefficient $A_K$ slowly decreases when $K$ increases. For instance, $A_5=0.872\dots$, 
$A_{10}=0.624\dots$. In the limit of large $K$, 
$A_K$ becomes zero. This comes from  the 
resummation
\begin{equation}
\sum_{l=1}^{\infty}\frac{\Gamma(l-1/2)}{l!}=\int_0^{\infty}dt t^{-3/2}(1-{\rm e}^{-t})=2\pi^{1/2}\ ,
\label{eq:gamma}
\end{equation}
which is exactly the $l=0$ term. In practice, when the order is not too large, 
$\beta^{-1}{\rm e}^{-2\beta}$ is a good order of magnitude estimate for the 
envelope discussed above as can be seen in Fig. \ref{fig:envelope}.

\section{Modified model with an arbitrary field cut}
\label{sec:cut}
In this section,we consider a modified partition function $Z(\beta, t_{max})$ where the integration range 
of Eq. (\ref{eq:tint}) takes a fixed, $\beta$-independent, value 
$t_{max}$. 
When $t_{max}<2\beta$, the Taylor series for the square root converges absolutely and 
uniformly over the whole range of integration. It is thus justified to interchange 
the sum and the integral and we have 
\begin{equation}
\label{eq:mod}
	Z(\beta, t_{max})=(\beta\pi)^{-3/2} 2^{1/2} 
	\sum_{l=0}^{\infty} A_l(t_{max})\beta^{-l}\ .
\end{equation}
The original partition function as expressed in Eq. (\ref{eq:notail}) is obtained by 
taking the limit $t_{max} \rightarrow 2\beta$. 
Since the integral with upper boundary $t_{max}$ is obviously continuous in that limit and since we can use the $l^{-2}$ suppression to prove the continuity of the sum, 
the validity of Eq. (\ref{eq:mod}) extends which justifies 
Eq. (\ref{eq:notail}). If $t_{max}>2\beta$, the sum diverges and the integral is ill-defined.
The regular perturbation series is obtained by taking the 
limit $t_{max}\rightarrow\infty$. In the graphs, we use the notation ``WK'' for the truncation of the regular perturbative series at order $K$. One should however keep in mind that,  
for instance in $W7$, the last term is of order $(1/\beta)^{7+3/2}$ due to the 
prefactor $\beta^{-3/2}$ in Eq. (\ref{eq:mod}). 

Eq. (\ref{eq:mod}) is now a regular series in $(1/\beta)$. It has a finite radius 
of convergence. In order to calculate this radius, we notice that 
for large $l$, $\int _0^{t_{max}}dt t^l $ gets most of its contribution from the 
region between $t_{max}(1-1/l)$ and $t_{max}$. Consequently, one can replace 
${\rm e}^{-t}t^{1/2}$ by ${\rm e}^{-t_{max}}t_{max}^{1/2}$ without affecting the 
asymptotic behavior of the coefficients of the series. 
If we perform this change directly in the integral Eq. (\ref{eq:tint}), 
the integral can be calculated explicitly. One can then conclude that 
$Z(\beta, t_{max})$ has a non-analytical part 
proportional to $(1-(t_{max}/2\beta))^{3/2}$. The series defined by Eq. (\ref{eq:mod}) converges if 
$(1/\beta)\leq (2/t_{max})$. Numerical studies of the series with conventional estimators confirm 
this argument.
Note that the finite radius of convergence of the series 
Eq. (ref{eq:mod}) is not in 
disagreement with the discontinuity of the original model at $1/\beta =0$, 
because this series coincides with the original model only when $(1/\beta)=(2/t_{max})$.

\begin{figure}[ht]
\includegraphics[width=3.4in,angle=0]{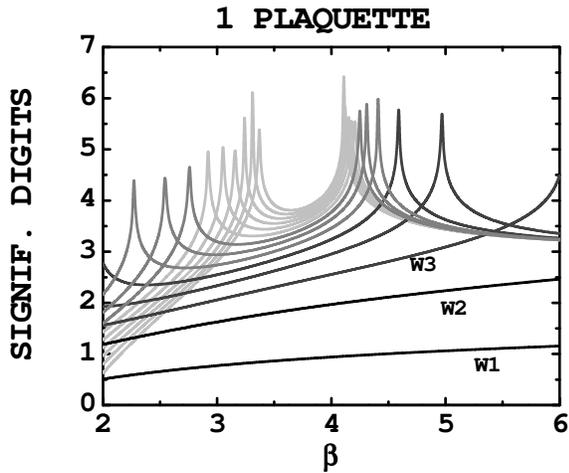}
\caption{Number of correct significant digits as a function of $\beta$  for a fixed cut $t_{max}=8$. As the 
order increases from 1 to 15 ($W1,\ W2,\dots$), the curves become lighter.
\label{fig:intg2}}
\end{figure}

Can a truncation of the series of Eq. (\ref{eq:mod}) at order $K$ be a good approximation of the original integral Eq. (\ref{eq:zsu2})? The answer depends on $K$, $t_{max}$ and 
$\beta$. It is clear that if $K$ is large enough and $\beta$ slightly above  $2/t_{max}$, 
then one should get a reasonable approximation. This statement is confirmed by Fig. \ref{fig:intg2} where the accuracy of 
Eq. (\ref{eq:mod}) with $t_{max}=8$ truncated at orders 1 to 15 is displayed as a function of $\beta$. In this particular case, the values 
of $\beta\geq 4$ are within the radius of convergence. As the order increases, the spikes in this region (the right half of the Fig. \ref{fig:intg2}) move toward 4. In addition, there is 
another set of spikes, outside the radius of convergence 
(on the left half of the Fig. \ref{fig:intg2}) and moving in the opposite 
direction when the order increases.

\begin{figure}[ht]
\includegraphics[width=3.4in,angle=0]{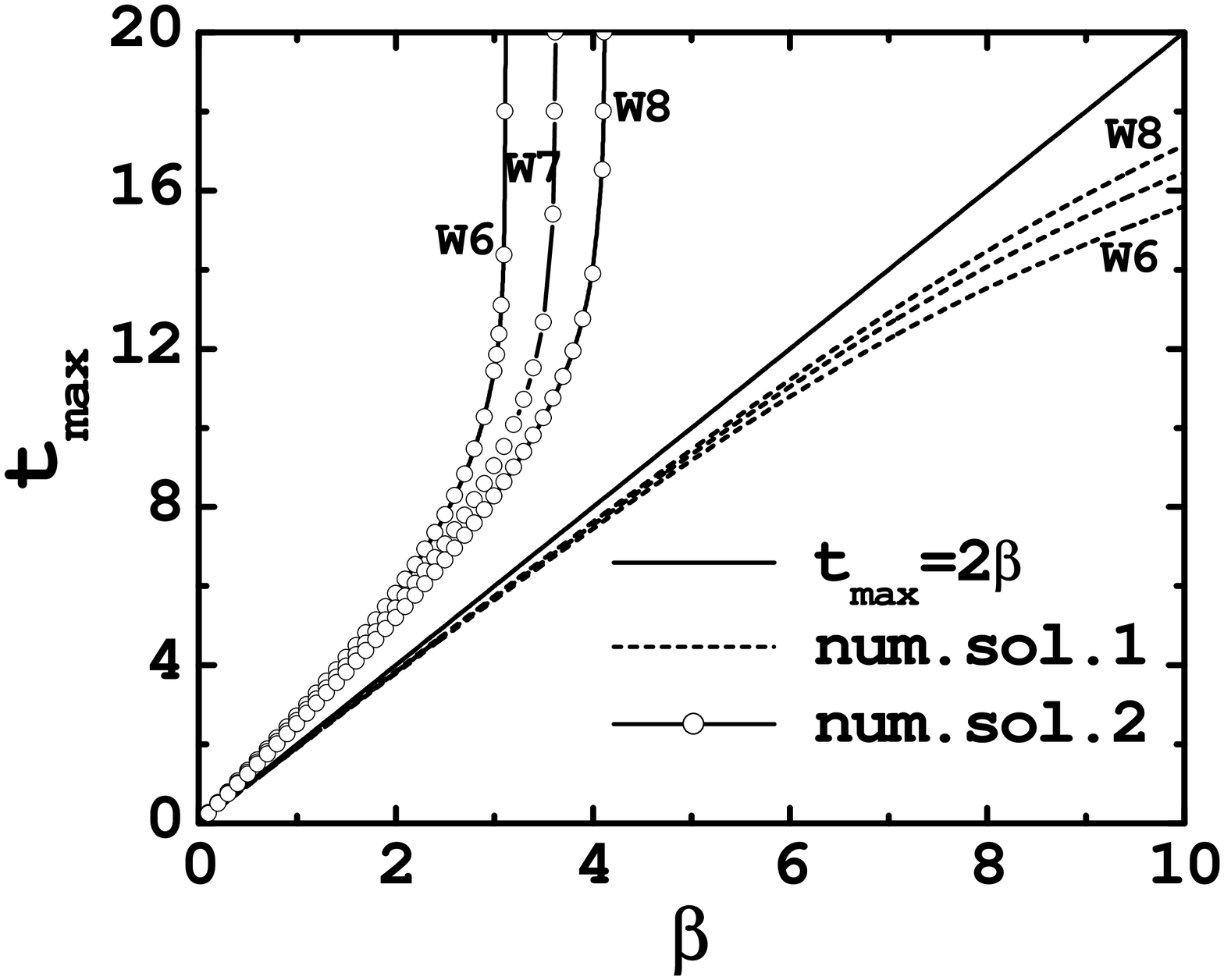}
\caption{ Location of the exact matching between the series Eq. (\ref{eq:mod}) at 
order 6, 7 and 8 and $Z(\beta)$ in the $\beta$-$t_{max}$ plane. The dashed lines 
represent the solution within the radius of convergence and the empty circles the other 
solution.
\label{fig:2sol}}
\end{figure}

A more global information regarding the location of the spikes is displayed in Fig. \ref{fig:2sol}.  It shows that the ``second solution'',  outside the radius of 
convergence, disappears beyond some critical value of $\beta$. As the order in the weak 
coupling increases, both solutions get closer to the $t_{max}=2\beta$ line.

\section{Optimization}
\label{sec:optim}
In this section, we discuss an approximate method to find the optimal value of $t_{max}$ corresponding to a given order $K$ and a given value of $\beta$. 
In a general situation, we do not know accurately the value 
of the quantity that we are calculating (the equivalent of $Z$ here).
Consequently, we need to find an approximation that allows us to 
consistently estimate this quantity and the way its order $K$ approximation 
changes with the field cutoff in order to impose an approximate 
matching condition.
For this purpose, we will use the strong coupling expansion (power series in 
$\beta$) which provides information complementary to the weak coupling. 
Now, the crucial point is that the field cut allows us to control the $(1/\beta)$
in the integral Eq. (\ref{eq:tint}), because (except in the exponential) all the factor $(1/\beta)$ 
appear together with a factor $t$. 
In other words, except for the exponential,it is a function of $t/\beta$.

\begin{figure}
\includegraphics[width=3.4in,angle=0]{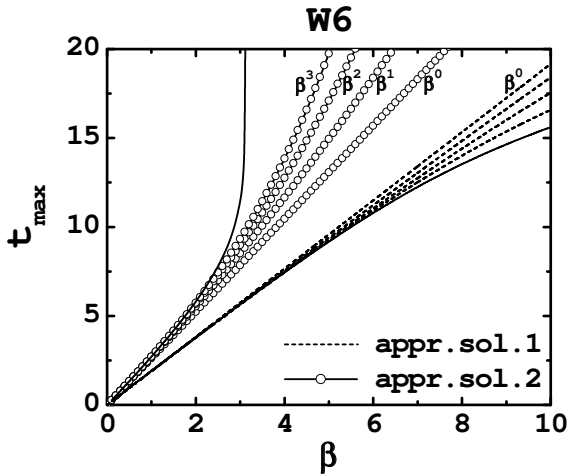}
\caption{Approximate locations in the $(\beta,t_{max})$ plane of the matching between the order 6 weak coupling 
expansion and $Z(\beta)$. The two solid lines are the two numerical solutions 
at that order (as in Fig. \ref{fig:2sol}). The dash line (empty circles) represent the first (second) approximate solutions at order $0,\dots,4$ in $\beta$.
\label{fig:appropt}}
\end{figure}

\begin{figure}[ht]
\includegraphics[width=3.4in,angle=0]{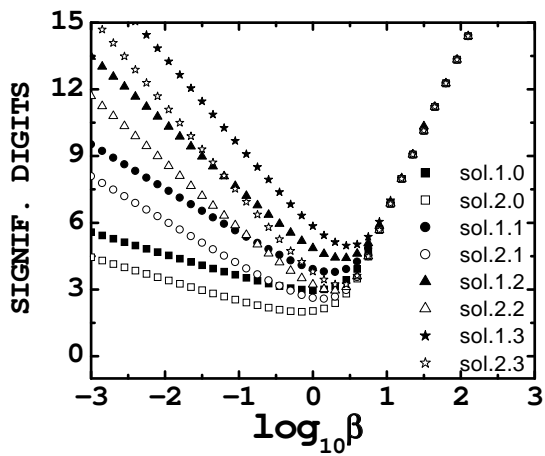}
\caption{Significant digits obtained from Eq. (\ref{eq:mod}) truncated at order 6 using 
$t_{max}/\beta$ at order 0 (squares), 1 (circles), 2 (triangles) and 3 (stars). 
The first solution with $t_{max}<2\beta$ is showed with filled symbols, while the 
second solution is showed with empty symbols.
\label{fig:sd2}}
\end{figure}

We would like to match the the strong coupling 
expansion 
\begin{equation}
Z(\beta)=1-\beta+(5/8)\beta^2 +\dots
\end{equation}
discussed in Sec. \ref{sec:model}, with
the truncated expansion of Eq. (\ref{eq:mod}) which can be rewritten as
	\[ \pi^{-3/2}2\sum_{l=0}^K\frac{\Gamma(l+1/2)}{l!(1/2-l)}\int_0^{t_{max}/\beta}ds 
	{\rm e}^{-s\beta}(s/2)^{l+1/2}\ .\]
The control of $s=t/\beta$ can be achieved by imposing that $t_{max}/\beta$ is approximately constant. We can then improve order by order in $\beta$ by setting
\begin{equation}
(t_{max}/\beta)=c_0 (K)+c_1(K)\beta +\dots
\label{eq:apbet}
\end{equation} 
The only non-trivial  part is to solve the zeroth order (in $\beta$) equation 
\begin{equation}
F_K(c_0 (K))=1\ ,
\label{eq:fk}
\end{equation}
 with 
\begin{equation}
F_K(x)=-4(\pi)^{-3/2}\sum_{l=0}^K\frac{\Gamma(l-1/2)	(x/2)^{l+3/2}}{l!(l+3/2)}
\ .
\end{equation}
We have checked that for $K$ going from 1 to 40,  Eq. (\ref{eq:fk}) has exactly two solutions on the positive real axis with one solution on each side of 2. As $K$ increases, the 2 roots get closer.  
They should coalesce at 2 in the large $K$ limit. This follows from the fact that  $F_{\infty}(2)=1$ 
and $F'_{\infty}(2)=0$ (as can be shown by using the the same method as for Eq. 
(\ref{eq:gamma})).
The higher order coefficients $c_l(K)$ corresponding to each 
of the two solutions at order 0 can then be found by solving linear equations. 
This procedure provides an approximate value of the optimal $t_{max}$ which apparently 
converges toward the correct numerical value. This is illustrated in Fig. \ref{fig:appropt}. 
If we plug the two approximate values of $t_{max}$ of Eq. (\ref{eq:apbet}) in Eq. (\ref{eq:mod}) truncated at order $K$, we obtain 
approximate values of $Z(\beta)$. The accuracy of this procedure is displayed in 
Fig. \ref{fig:sd2} in the case $K=6$. It appears clearly that the first solution 
(the one within the radius of convergence with $t_{max}<2\beta$) is significantly more accurate than the 
other solution (with $t_{max}>2\beta$). Similar features were observed for $K$ up to 20. 

\begin{figure}[ht]
\includegraphics[width=3.4in,angle=0]{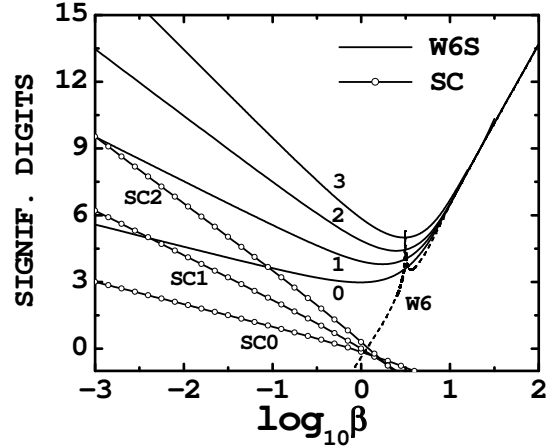}
\caption{ Significant digits obtained from Eq. (\ref{eq:mod}) truncated at order 6 using the first solution for 
$t_{max}/\beta$ at order 0 to 3 compared to the weak coupling expansion at order 6 
(dotted line W6) and the strong coupling expansion at order 0 to 2 (empty circles SC)
\label{fig:sd}}
\end{figure}

We can now compare the accuracy of the method proposed here with the weak and strong 
coupling expansions. The case $K=6$ is displayed in Fig. (\ref{fig:sd}). In the weak 
coupling region ($\beta>10$) the accuracy of our procedure merges with the regular 
perturbation series. In the strong coupling region ($\beta<0.1$), our procedure 
is more accurate than the regular expansion in powers of $\beta$ by several 
significant digits. As $\beta\rightarrow 0$, the accuracy of our procedure 
with $t_{max}/\beta$ determined at order $m$ in $\beta$ increases at the 
same rate as the regular strong coupling expansion at order $m$ in $\beta$ maintaining 
the difference in accuracy approximately constant. In the intermediate region where 
none of the conventional expansions work well (except at the perturbative spike), our procedure 
maintains a very good accuracy interpolating smoothly between the two regimes.

\section{Asymptotic behavior of $c_l(K)$}
\label{sec:asy}

\begin{figure}[ht]
\includegraphics[width=3.4in,angle=0]{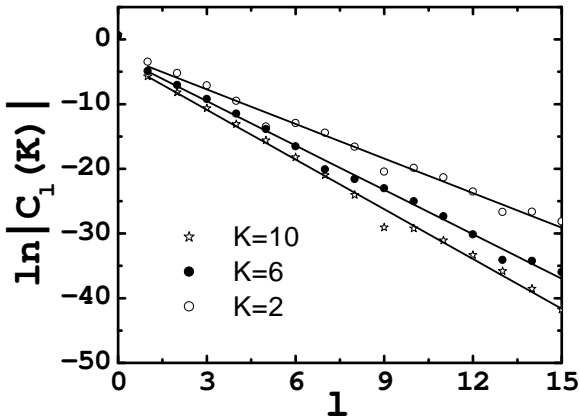}
\caption{ $\ln |c_l(K)|$ versus $l$ for $K=2, \ 6\ ,10$.
\label{fig:ck1}}
\end{figure}

In this section, we study empirically the asymptotic behavior of the coefficients $c_l(K)$ appearing in the expansion of $t_{max}/\beta$ Eq. (\ref{eq:apbet}). At fixed $K$ large $l$ , 
Fig. \ref{fig:ck1} suggests that 
\begin{equation}
c_l(K)\propto(G(K))^l	\ .
\end{equation}
In addition, it appears that $G(K)$ decreases with $K$ approximately like $1/K$.
This behavior implies a finite radius convergence $G(K)^{-1}$ for the $\beta $ expansion in Eq. (\ref{eq:apbet}), increasing linearly with $K$. This is good news 
for the interpolation between the weak and strong coupling region since
as we increase the weak order $K$, we increase the range of validity in $\beta$.

\begin{figure}[ht]
\includegraphics[width=3.4in,angle=0]{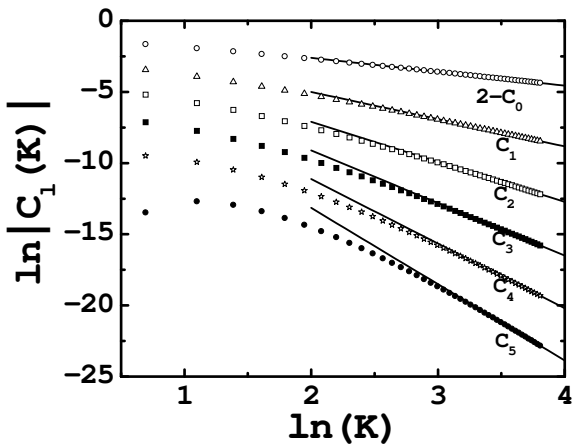}
\caption{$\ln |c_0(K)-2|$ versus $\ln K$; $\ln |c_l(K)|$ versus $\ln K$ for $l=1,\dots 5$.
\label{fig:ck2}}
\end{figure}

The large-$K$ behavior of $c_l(K)$ has also been studied numerically. 
The results for $l$ up to 5 are shown in Fig. \ref{fig:ck2}.
At fixed $l$ large $K$ , the linear fits used in Fig. \ref{fig:ck2} suggest that
\begin{equation}
c_0 (K)\simeq 2+O(1/K)	\ , 
\end{equation}
and 
\begin{equation}
c_l(K)\propto K^{-l-1+\alpha(l)}	\ , 
\end{equation}
for $l>0$, with $\alpha(l)$ small.
This behavior is expected, since as the order 
increases, we are getting close to the
exact expansion Eq. (\ref{eq:notail}) with $t_{max}=2\beta$ ($c_0=2,\ c_l=0$ for $l>0$).
The values of $\alpha(l)$ decrease when we reduce the set of points fitted to
larger values of $K$. If we use $K=35$ to 45 for the fit, we have approximately
$\alpha(l)\simeq l/10$.

\section{Conclusions}
We have shown that for the one-plaquette model, the introduction of a properly chosen
field cut can provide a high accuracy
in regions where the usual perturbative method is not accurate.
The strong coupling expansion provides an efficient way to determine the optimal cut
and interpolate between the small and the large $\beta$ regions. 
Apparently, the 
accuracy of the calculations improves whenever we increase the order in either 
the weak or the strong coupling expansion. Given these positive results, we are compelled 
to implement the method in the case of lattice gauge theory 
on a $D$ dimensional cubic lattice. Two steps are necessary.
First, we need to define the theory with a field cut (the analog of Eq. (\ref{eq:tint}) with $2\beta$ replaced by $t_{max}$). Second, we need to 
expand relevant quantities such as $P$ for the modified theory in powers 
of $1/\beta$ (the analog of Eq. (\ref{eq:mod})). 
Note that in the calculation of P using a perturbative series, 
the complex zeroes of Z will play an important role. This question remains to 
be examined in detail.

The implementation of the first step is straightforward. One can insert 
1 in the partition function in the following way:
\begin{equation}
1=\prod_p\int_0^{t_{max}}dt_p\delta(1-(1/N)ReTr(U_p)-t_p)\ .
\end{equation}
If we could perform the integration over the $U_{link}$, we would get an effective theory 
for the new variables $t_P$. Note that the procedure is gauge invariant since 
$TrU_p$ is. The ``size'' of a configuration can be defined in several ways.
For instance, we could use the value of $ Max_p \left\{ t_p \right\} $ or $(1/N_p)\sum_p t_p$ to decide
if we have a large or a small field configuration. We can then order the configurations according to the chosen indicator. Given a (sufficiently large) set 
of Monte Carlo configurations, one can define the expectation values with a cut 
by averaging only over configurations for which the chosen indicator is below 
a certain value. The correlations between the two size indicators mentioned 
above are now being studied for $SU(3)$ in 4 dimensions.

The implementation of the second step requires technologies that are now being 
developed in the scalar case. As it seems only possible to make analytical 
calculations for small or large field cuts, numerical methods seem unavoidable.
For the purpose of independent verification, it is important to consider different 
methods. We are presently working on three different approaches:
\begin{enumerate}
\item
The conventional 
approach \cite{heller84} but with the  $A_\mu^a$ having a finite range of integration.
This type of approach works well in the scalar case \cite{lilipro}
\item
The stochastic approach \cite{direnzo95} where $A_\mu^a$ is expanded as 
power series in $1/\beta$. For the lowest order field, the implementation 
of a cut is obvious but not for higher order fields. This problem is being considered 
with simple examples.
\item
Fits from numerical data at large $\beta$. This method \cite{cookpro} 
allowed to extract at least 
2 coefficients of conventional perturbation. As we mentioned above, it is easy to 
implement the field cut with Monte Carlo methods. The advantage of this method is
that it does not require the use of the Campbell-Baker-Hausdorff (for a short review see Ref. \cite{cbh}) formula.
\end{enumerate}
We expect that the use of theses three methods will allow us to 
contruct perturbative series with a finite radius of convergence as above. 
We hope that this radius of convergence will be sufficiently large to 
reach the scaling window. Ultimately, we expect to be able to replace the numerical calculation of the 
coefficients by approximate analytical formulas, as it seems possible to do 
in the case of the anharmonic oscillator \cite{interp}.

\begin{acknowledgments}
This research was supported in part by the Department of Energy
under Contract No. FG02-91ER40664.  
We thank Antonio Gonzalez Arroyo for discussions on Bessel functions.
\end{acknowledgments}

%\bibliography{c:/papers/mainbib}

\end{document}